\documentclass[12pt,preprint]{aastex}
\slugcomment{Accepted for publication in the ApJL on July 27th, 2011}
\shorttitle{Red Nuggets at High Redshift}
\shortauthors{Damjanov et al.}
\begin{document}
\title{Red Nuggets at High Redshift: Structural Evolution of Quiescent Galaxies Over 10~Gyr of Cosmic History}

\author{
Ivana Damjanov\altaffilmark{1}, 
Roberto G. Abraham\altaffilmark{1}, 
Karl Glazebrook\altaffilmark{2}, 
Patrick J. McCarthy\altaffilmark{3}, 
Evelyn Caris\altaffilmark{2},
Raymond G. Carlberg\altaffilmark{1},
Hsiao-Wen Chen\altaffilmark{4},
David Crampton\altaffilmark{5}, 
Andrew W. Green\altaffilmark{2},
Inger J{\o}rgensen\altaffilmark{6}, 
St{\'e}phanie Juneau\altaffilmark{7}, 
Damien Le Borgne\altaffilmark{8}, 
Ronald O. Marzke\altaffilmark{9},
Erin Mentuch\altaffilmark{1,10},
Richard Murowinski\altaffilmark{5}, 
Kathy Roth\altaffilmark{6}, 
Sandra Savaglio\altaffilmark{11}, 
Haojing Yan\altaffilmark{12}
}

\altaffiltext{1}{Department of Astronomy \& Astrophysics, University of Toronto, 50 St. George Street, Toronto, ON, M5S~3H4}
\altaffiltext{2}{Centre for Astrophysics and Supercomputing, Swinburne University of Technology, 1 Alfred St, Hawthorn, Victoria 3122, Australia}
\altaffiltext{3}{Observatories of the Carnegie Institution of Washington, 813 Santa Barbara Street, Pasadena, CA 91101}
\altaffiltext{4}{The Department of Astronomy and Astrophysics, University of Chicago, 5640 S. Ellis Ave, Chicago, IL 60637}
\altaffiltext{5}{Herzberg Institute of Astrophysics, National Research Council, 5071 West Saanich Road, Victoria, British Columbia, V9E~2E7, Canada} 
\altaffiltext{6}{Gemini Observatory, Hilo, HI 96720}
\altaffiltext{7}{Department of Astronomy/Steward Observatory,University of Arizona,933 N Cherry Ave., Rm. N204,Tucson AZ 85721-0065}
\altaffiltext{8}{Institut d'Astrophysique de Paris, UMR 7095, CNRS, UPMC Univ. Paris 06, 98bis boulevard Arago, F-75014 Paris, France}
\altaffiltext{9}{Dept. of Physics and Astronomy, San Francisco State University, 1600 Holloway Avenue, San Francisco, CA 94132}
\altaffiltext{10}{Department of Physics \& Astronomy, McMaster University, 1280 Main St. W, Hamilton, ON L8S 4M1 Canada}
\altaffiltext{11}{Max-Planck-Institut f\"ur extraterrestrische Physik, Garching, Germany}
\altaffiltext{12}{Center for Cosmology \& AstroParticle Physics, The Ohio State University, 191 West Woodruff Ave, Columbus, OH 43201}

\defcitealias{abr04}{Paper~I}
\defcitealias{gla04}{Paper~III}
\defcitealias{mcc04}{Paper~IV}
\defcitealias{abr07}{Paper~VIII}

\begin{abstract}

\medskip
We present an analysis of the  size growth seen in early-type galaxies over 10 Gyr of cosmic time.
Our analysis is based on a homogeneous synthesis of published data from 17 spectroscopic surveys observed at similar spatial resolution, augmented
by new measurements for galaxies in the Gemini Deep Deep Survey.
In total, our sample contains structural data for 465
galaxies (mainly early-type)  in the redshift range $0.2<z<2.7$. 
The size evolution of passively-evolving galaxies over this redshift range
is gradual and continuous, with no evidence for an end or change to the process around
$z\sim1$, as has been hinted at by some surveys which analyze subsets of the data
in isolation. The size growth appears to be 
independent of stellar mass, with the mass-normalized half-light radius
scaling with redshift as $R_e\propto(1+z)^{-1.62\pm0.34}$.  Surprisingly, this power law seems to be in good agreement with the recently 
reported continuous size evolution of UV-bright galaxies in the redshift range $z\sim0.5-3.5$. It is also in accordance with the predictions from recent theoretical models. 

\medskip

{\emph{Subject headings:}\rm{ galaxies:~elliptical, galaxies:~fundamental parameters, galaxies:~evolution}}

\medskip

{\emph{ On-line material:}\rm{ machine-readable table}}  

\end{abstract}


\section{Introduction}

The discovery of a puzzling new population of compact $(R_e\lesssim1\, \mathrm{kpc})$ massive elliptical galaxies existing at  epoch when the Universe was not more than one-third of its current age has posed profound challenges for both monolithic and hierarchical model of galaxy formation and evolution. A handful of these objects were first reported by \citet{Cimatti2004}, and later work by several groups
has grown the number of similar galaxies at redshifts $z\gtrsim1.5$ by more than a factor of 30 \citep[e.g,][]{Daddi2005,Trujillo2006,Longhetti2007,Toft2007,Zirm2007,Buitrago2008,Cimatti2008,VanDokkum2008,Damjanov2009,Cassata2010,Newman2010,Ryan2010,Saracco2010}. Although some concerns have been noted regarding the uncertainties in size measurements based on the {\em Hubble Space Telescope} (HST) imaging \citep{Mancini2010}, recent results based on the ultra-deep HST WFC3 data \citep{Cassata2010,Ryan2010,Szomoru2010} have confirmed that the typical sizes of quiescent galaxies at high redshifts are several times smaller that the sizes of their local massive counterparts. Furthermore, visible and near-infared (NIR) spectroscopy of individual high-$z$ `red and dead' galaxies have revealed high velocity dispersions and central stellar densities \citep{VanderWel2008,vanDokkum2009,Newman2010,VandeSande2011} which are consistent with those expected from compact galaxies.

In the present paper we synthesize the results from these published surveys, which together span a redshift range from the nearby Universe ($z\sim0.2$) all the way out  to redshifts $z\sim2.7$.  This redshift range spans $\sim10~$Gyr of cosmic time.  By combining  published data with new measurements for galaxies in the Gemini Deep Deep Survey (GDDS), we are able to compile a sample of 465 galaxies with spectroscopic redshifts over the full redshift range. Our
main aim is to use these galaxies to  determine whether galactic size growth is a continuous process that occurs over this full redshift range, or a process that is mainly associated with a particular epoch. Our result will place additional constraints on two mechanisms that have been proposed to explain the observed size growth:  (1) minor dry mergers or late accretion \citep[e.g.,][]{Oser2011} and (2) adiabatic expansion due to extreme mass loss \citep[caused by stellar winds or quasar activity,][]{Damjanov2009,Fan2010}.

\section{Sample and Data Reduction}\label{data}

\begin{figure*}[htp!]
\begin{center}
\includegraphics[scale=.6,angle=90]{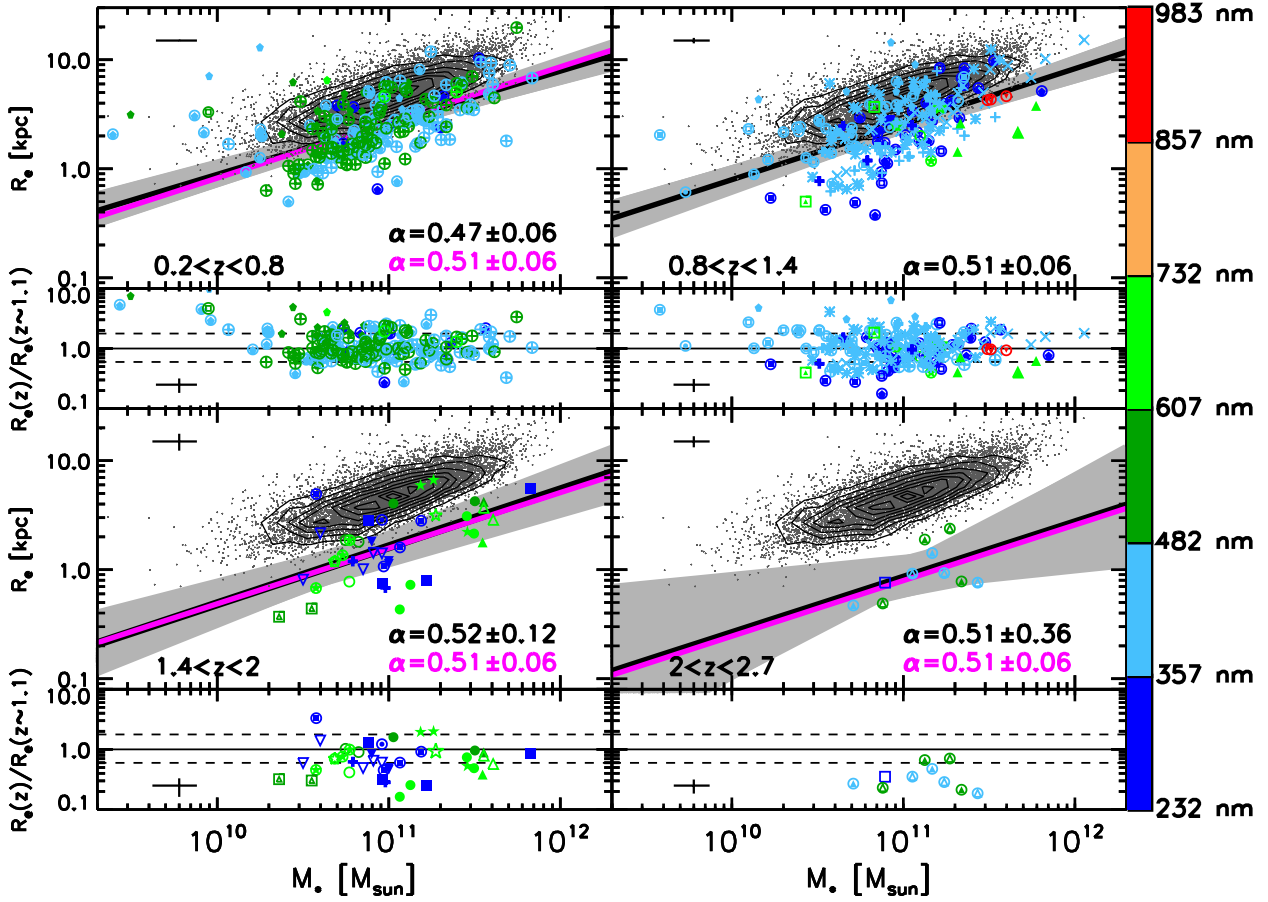}
\vspace {1cm}
\caption{Effective radius $\rm{R}_e$ as a function of stellar mass for a sample of $465$ passively evolving galaxies in the redshift range $0.2<z<2.7$. Different symbols correspond to different surveys (listed in the legend of Figure~\ref{f3}) and are color coded based on the \it{rest-frame}\rm \ central wavelength of the size measurements with the key shown as a color bar at right. Data points are compared to the local sample drawn from the SDSS (grey points) in separate redshift panels. Contours represent linearly spaced regions of constant density of local SDSS galaxies in size-mass parameter space. The solid black line and grey area represent the best-fit relation to the data points in each redshift bin and its $\pm1\sigma$~errors, respectively. In each upper sub-panel the slope $\alpha$ of the magenta line is the best fit to the data in a given redshift range with the slope fixed to the slope of the $0.8<z<1.4$ relation. (Note that the linear fits exclude objects with masses $<10^{10}M_{\odot}$ to avoid being skewed by very low-mass outliers.). Average error bars for objects in different redshift bins are given in the left top corner of each panel. Note that we do not have information on the size measurement errors for $>95\%$ of objects at $z<1$ (Table~\ref{tab2}). Lower sub-panels show the ratio between the measured size and the size at $0.8<z<1.4$ based on the size-mass relation plotted in the upper panels, as a function of mass. The solid line corresponds to the same sizes in a given redshift bin and at $0.8<z<1.4$ $(R_e(z)/R_e(z\sim1.1)=1)$, and the dotted lines encompass the $\pm 1 \sigma$ spread of the $z\sim1.1$ data. \label{f1}}
\end{center}
\end{figure*}

\begin{figure*}[htp!]
\begin{center}
\vspace {1cm}
\includegraphics[scale=.6,angle=90]{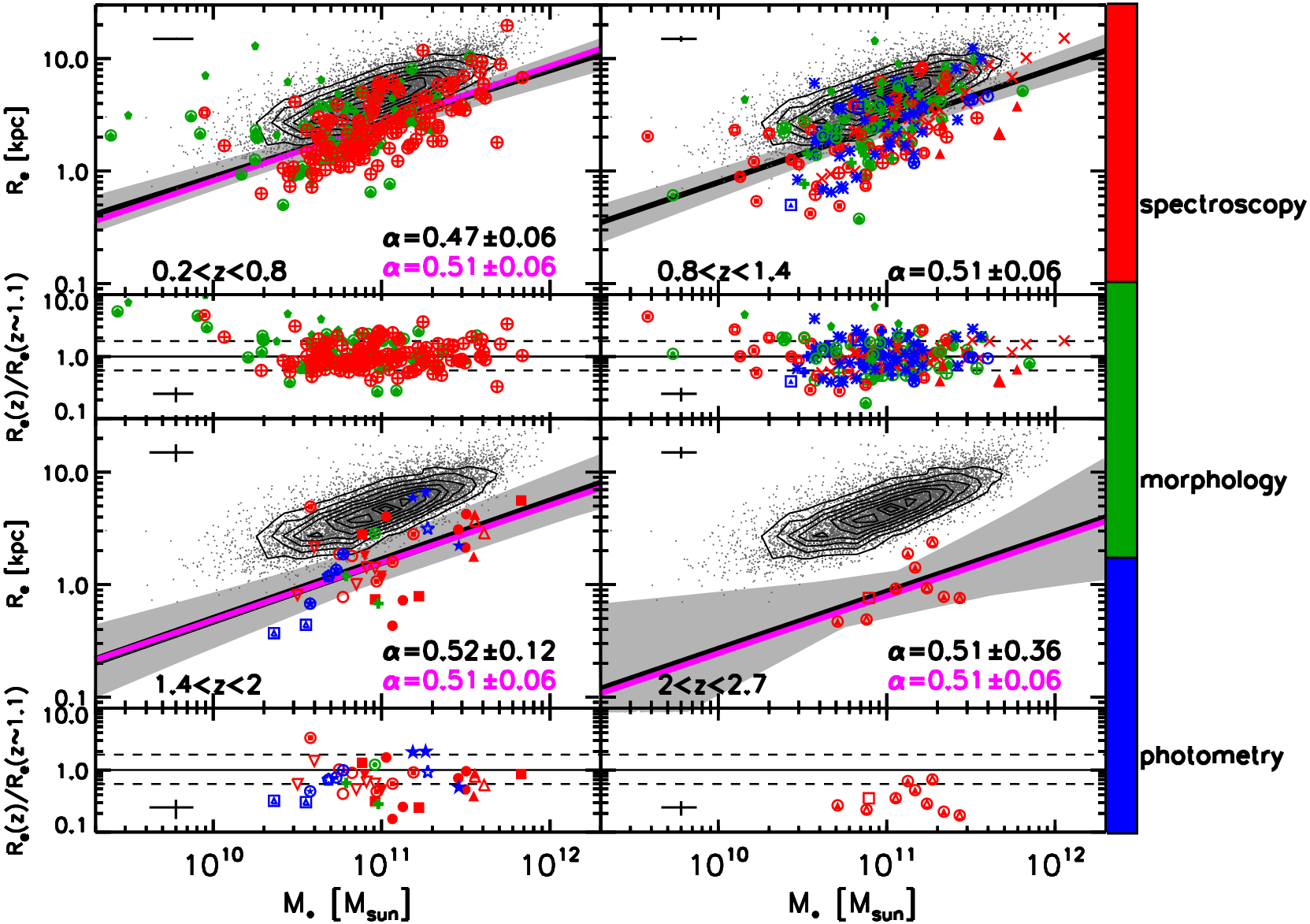}
\vspace {1cm}
\caption{Effective radius $\rm{R}_e$ as a function of stellar mass for a sample of $465$ passively evolving galaxies in the redshift range $0.2<z<2.7$. The notation is the same as in Figure~\ref{f1} except that the  color coding is now based on the sample selection criteria given in Table~\ref{tab1}. \label{f2}}
\end{center}
\end{figure*}

Table~\ref{tab1} presents a summary of the the structural parameters based on high resolution HST and adaptive optics ground-based imaging for 434 galaxies obtained from the literature for
16 spectroscopic surveys, augmented with additional analysis of imaging data for 31 objects from our own survey \citep[GDDS;][]{Abraham2004}. 
The available data include redshifts, stellar masses, and the S\'ersic surface brightness profile parameters - circularized half-light radii $R_e$ and S\'ersic profile indices $n$ - in the rest frame. All data were harmonized to a common
cosmology (H$_0=70$~km/s/Mpc, $\Omega_\textrm{m}=0.3$, $\Omega_\Lambda=0.7$). Likewise, stellar masses were harmonized to a common initial mass function \citep[IMF,][]{Baldry2003}.
In six surveys indices $n$ were not available for individual objects : EDisCS, CFRS, GN/DEIMOS, MS1054/CDFS, CL1252/CDFS, and EGS/SSA22/GN. Four of these surveys reported $R_e$ corresponding to the best fitting de Vaucoulers ($R^{1/4}$) profile for all objects. The median S\'ersic index of the EdisCS sample is $<n>=3.7$, and 19 objects in Figure~4 of \cite{Saglia2010} are best described with $n\lesssim2.5$ profiles. The CL1252/CDFS survey  presented in \citet{Rettura2010} provided S\'ersic profiles of quiescent galaxies along with their half-light radii, but without corresponding  S\'ersic indices. In eight out of 11 remaining surveys, for which the shape of galaxy's surface brightness profiles can be classified as disk-like or spheroid-like based on the S\'ersic index, the majority of quiescent galaxies and compact objects are spheroids with $n\geqslant2.5$. The three spectroscopic surveys in our compilation with more than 50\% of disk-dominated compact objects are the MUSYC survey \citep{VanDokkum2008}, the survey presented in \citet{Cassata2010}, and the GMASS \citep{Cimatti2008}. While the MUSYC focuses on galaxies at $z>2$, 67\% (4/6) of 
disk-like objects with compact morphologies from \citet{Cassata2010} and \citet{Cimatti2008} are found at $z\sim1.9$. We note that \citet{VanderWel2011} claim that the majority of massive compact galaxies at $z\sim2$ are disk-dominated.   

In compiling the data summarized in Table~\ref{tab1} we augmented our published
NICMOS sizes for quiescent galaxies in the GDDS fields \citep{Damjanov2009} with
additional measurments obtained from imaging with the Advanced Camera for Surveys (ACS). The
observational strategy for this ACS imaging was laid out in \citet{Abraham2007}, and the mages were processed using the  technques described in 
 \cite{Damjanov2009}. Out of the 40 quiescent GDDS galaxies imaged with ACS, four objects were  also imaged with HST NICMOS Camera 3 in $H$~band in \cite{Damjanov2009}. Sizes obtained with ACS
agreed to within $\lesssim25\%$ with those obtained from NICMOS. In cases with duplicate measurments 
we chose to retain the NICMOS sizes because
they probe longer rest frame wavelengths, which are less affected by dust extinction and are better
tracers of total stellar emission. Residual images for another three galaxies show asymmetric features and their 2D profiles cannot be reliably modelled with one-component or two-component S\'ersic profiles. Finally, two objects appear 
in the ACS images as mergers (one merging pair is spectroscopically confirmed) and modelling of their 2D profiles depends critically on details
of how the companion galaxy is masked, so we omitted these galaxies from our sample. After making these cuts,
an additional 31 GDDS quiescent massive galaxies were added to our total sample.

The complete list of objects in our compilation with all their properties we used to construct relations presented in this paper is given in Table~\ref{tab2}. We note that there are overlaps between a few $z>1$ samples drawn from the  south field of the Great Observatories Origin Deep Survey (GOODS): MS1054/CDFS, CL1252/CDFS, GS/WFC3, GS/ACS, HUDF/WFC3, GMASS, and HUDF. In order to exclude all duplicate entries for the objects with unpublished positions we flagged all galaxies in Table~\ref{tab2} having the same redshifts and similar mass and size estimates and kept the results based on deeper imaging (e.g., WFC3) whenever possible. Our approach ensures that all 465 entries in Table~\ref{tab2} are unique.

\begin{deluxetable}{lcccccccccc}
\tablecolumns{11}
\tablewidth{0pc}
\tablecaption{Summary of the compilation of samples used to construct the size evolution diagram\label{tab1}}
\tabletypesize{\tiny}
\tablehead{
\colhead{Sample\tablenotemark{a}} & \colhead{$z_{\mathrm{spec}}$} & \colhead{$\lambda_\mathrm{rest}(R_e)$} & \colhead{$M_\ast$\tablenotemark{b}} & \colhead{N} & \colhead{$n\geqslant2.5$}  & \colhead{Quiescent }  & \colhead{$n\geqslant2.5$} & \colhead{Compact} & \colhead{$n\geqslant2.5$} & \colhead{Ref}\\
\colhead{} & \colhead{} & \colhead{} & \colhead{} & \colhead{} & \colhead{} & \colhead{} & \colhead{quiescent} &  \colhead{} & \colhead{compact} &\colhead{}  \\
\colhead{}& \colhead{} & \colhead{(nm)} &\colhead{$(10^{11}$~M$_\sun)$} &  \colhead{} & \colhead{(\%)} & \colhead{(\%)}& \colhead{(\%)}& \colhead{(\%)} & \colhead{(\%)} & \colhead{} \\
\colhead{(1)}& \colhead{(2)} &\colhead{(3)} &  \colhead{(4)} & \colhead{(5)} & \colhead{(6)}& \colhead{(7)}& \colhead{(8)} & \colhead{(9)} & \colhead{(10)} & \colhead{(11)} \\
}
\startdata
EDisCS\dotfill&      0.24-0.96   &  415- 656   &    0.12- 6.85 &  154 &  87.66 & 100.00 &  87.66 &  23.37 &  $\geqslant47.22$ &    1  \\
\bf{CFRS}\rm\dotfill&      0.29-0.99  &  409- 631 &  0.04- 3.09 &      36 & 100.00 &  72.50 & 100.00 &   5.55 & 100.00 &    2     \\
\bf{GN/DEIMOS}\rm\dotfill&      0.18-1.14 &  283- 514 &  0.03- 7.04 &      76 & 100.00 &  75.00 & 100.00 &  26.32 & 100.00 &   3,4     \\
MS1054/CDFS\dotfill&     0.84-1.14 &  353- 464 &  0.42-11.33\tablenotemark{c} &      32 & 100.00 & 100.00 & 100.00 &   9.37 & 100.00 &    5    \\
\it{CL1252/CDFS}\rm\dotfill&       1.09-1.35 &  362- 407 &  0.29- 3.64  &      44 & N/A & 100.00 & N/A &  25.00 & N/A &    6   \\
\bf{EGS/SSA22/GN}\rm\dotfill&       1.05-1.59 &    328- 397   &0.33- 1.55 &      17 & 100.00 & 100.00 & 100.00 &  35.29 & 100.00 &    7  \\
\it{Radio-loud QSOs}\rm\dotfill&       1.29-1.59 &    618- 699  &1.54- 2.87 &       5 &  60.00 & 100.00 &  60.00 &   0.00 &   0.00 &   8,9   \\
MUNICS\dotfill&       1.23-1.71 &  590- 717  & 2.06- 5.95 &        9 &  66.66 & 100.00 &  66.66 &  11.12 & 100.00 &   10   \\
\it{GS/WFC3}\rm\dotfill&       1.33-1.62  &  611- 687  &  0.37- 1.45&       6 &  66.66 & 100.00 &  66.66 &  66.66 &  75.00 &   11   \\
GDDS/ACS\dotfill&      0.62-1.74 &    297- 502   &  0.04- 2.25 &    31 &  54.84 & 100.00 &  54.84 &  41.94 &  53.85 &   12  \\
\it{EGS}\rm\dotfill&      1.24-1.36 &    932- 982   &3.09- 3.98 &       3 &  66.66 &  N/A &  N/A&   0.00 &   0.00 &   13  \\
GDDS/NICMOS\dotfill&       1.39-1.85 &  561- 669   &  0.55- 3.17 &      10 &  60.00 &  90.00 &  55.55 &  30.00 &  66.66 &   14  \\
\bf{GS/ACS}\rm\dotfill &    0.95-1.92 &   291- 436 &  0.05- 2.08 &     15 & 100.00 & 100.00 & 100.00 &  13.34 & 100.00 & 15,16    \\
\it{HUDF/WFC3}\rm\dotfill&      1.32-1.98 & 537- 690  &    0.23- 0.67 &      4 &  50.00 & 100.00 &  50.00 &  75.00 &  33.34 &   17   \\
GMASS\dotfill&       1.42-1.98 &    285- 351   & 0.32- 0.99  &      8 &  37.51 & 100.00 &  37.51 &  75.00 &  33.34 &   18 \\
HUDF\dotfill&       1.39-2.67 &   232- 356   &  0.76- 6.74 &      6 &  83.34 & 100.00 &  83.34 &  50.00 &  66.66 &   19  \\
MUSYC\dotfill&      2.03-2.55 &    451- 528  &   0.52- 2.71  &    9 &  44.45 & 100.00 &  44.45 &  77.77 &  42.85 &   20  \\
\hline
\\
TOTAL\dotfill& 0.2 -- 2.67 & 232 -- 982 &0.03 -- 11.33& 465 & $\geqslant78.07$&$\geqslant92.90$ &$\geqslant76.09$&25.80& $\geqslant59.17$& \\ 
\enddata
\tablecomments{Column 1: survey from which the sample is drawn; Column 2: redshift range; Columns 3: the range of rest-frame central wavelengths of the $R_e$ measurements; Column 4: mass range; Column 5: number of objects in the sample; Column 6: fraction of passively evolving objects; Column 7: fraction of spheroids; Column 8: fraction of passively evolving galaxies with spheroid-like profiles; Column 9: fraction of  (compact) objects with $R_e\lesssim 1$~kpc; Column 10: fraction of compact objects with spheroid-like profiles ; Column 11: references: 1. \citet{Saglia2010}; 2. \citet{Schade1999}; 3. \citet{Treu2005}; 4. \citet{Bundy2007}; 5. \citet{VanderWel2008}; 6. \citet{Rettura2010}; 7. \citet{Newman2010}; 8. \citet{McGrath2007}; 9. \citet{McGrath2008}; 10. \citet{Longhetti2007}; 11. \citet{Ryan2010}; 12. data presented here; 13.\citet{Carrasco2010}; 14. \citet{Damjanov2009}; 15. \citet{Gargiulo2011}; 16. \citet{Saracco2010}; 17. \citet{Cassata2010}; 18. \citet{Cimatti2008}; 19. \citet{Daddi2005}; 20. \citet{VanDokkum2008}}
\tablenotetext{a}{Selection criteria for each sample are denoted by the font style: roman denotes spectroscopically selected objects with old stellar population, boldface is used for morphologically selected early-type galaxies, and italics font corresponds to the quiescent galaxies selected by colour.} 
\tablenotetext{b}{Stellar mass estimates have been converted to the \citet{Baldry2003}~IMF.}
\tablenotetext{c}{Based on dynamical masses $M_\mathrm{dyn}$ and the $M_\mathrm{dyn}\sim1.4\times M_*$ relation \citep{VanderWel2008}.}
\end{deluxetable}

\begin{deluxetable}{lccccccccccc}
\tablecolumns{12}
\rotate
\tablewidth{0pc}
\tablecaption{Complete list of objects used to construct the size evolution diagram \label{tab2}}
\tabletypesize{\tiny}
\tablehead{
\colhead{Object ID} & \colhead{R. A.} & \colhead{Dec} & \colhead{$z_{\mathrm{spec}}$}& \colhead{Selection} & \colhead{$M_\ast$} & \colhead{$\Delta M_\ast$} & \colhead{Observing filter}  & \colhead{R$_e$}  & \colhead{$\Delta \mathrm{R}_e$} & \colhead{$n$} & \colhead{$\Delta n$}\\
\colhead{} & \colhead{J2000 (deg)} & \colhead{J2000 (deg)} & \colhead{} & \colhead{} & \colhead{$(10^{11}$~M$_\sun)$} & \colhead{$(10^{11}$~M$_\sun)$} & \colhead{} &  \colhead{(kpc)} & \colhead{(kpc)} &\colhead{}  & \colhead{}\\
\colhead{(1)}& \colhead{(2)} &\colhead{(3)} &  \colhead{(4)} & \colhead{(5)} & \colhead{(6)}& \colhead{(7)}& \colhead{(8)} & \colhead{(9)} & \colhead{(10)} & \colhead{(11)} & \colhead{(12)}\\
}
\startdata
         EDCSNJ1040403-1156042 & 160.167917 & -11.934500 & 0.7020 &                   spectroscopy &   2.663 &   0.368 &  F814W &   6.153 & \dotfill &   3.700 &  \dotfill  \\
         EDCSNJ1040407-1156015 & 160.169583 & -11.933750 & 0.7030 &                   spectroscopy &   1.463 &   0.168 &  F814W &   1.698 &  \dotfill &   3.700 &  \dotfill  \\
         EDCSNJ1040346-1157566 & 160.144167 & -11.965722 & 0.7024 &                   spectroscopy &   1.366 &   0.252 &  F814W &   3.348 &  \dotfill &   3.700 &   \dotfill  \\
         EDCSNJ1040396-1155183 & 160.165000 & -11.921750 & 0.7046 &                   spectroscopy &   0.945 &   0.196 &  F814W &   2.244 &  \dotfill &   3.700 &   \dotfill   \\
         EDCSNJ1040356-1156026 & 160.148333 & -11.934056 & 0.7081 &                   spectroscopy &   1.719 &   0.475 &  F814W &   2.345 &  \dotfill &   3.700 &  \dotfill  \\
\enddata
\tablecomments{Table~\ref{tab2} is published in its entirety in the electronic edition of the ApJ Letters. A portion is shown here for guidance.}
\end{deluxetable}

\section{The size-mass relation}

Figure~\ref{f1} presents the size-mass relation obtained from nearly 500 massive galaxies with known structural parameters spanning the (spectroscopically confirmed) redshift range from $z_{\mathrm{spec}}\sim0.2$ to $z_{\mathrm{spec}}\sim2.7$.
The figure shows the data in six different redshift bins and in each panel  the high-redshift sample is shown relative to the
local distribution of galaxies on the size-mass plane. These local data are from the Sloan Digital Sky Survey (SDSS), with sizes taken from \citet{Bernardi2003} and matched with masses calculated following \citet{Baldry2008}. The linear relation
shown in each panel is the best-fit line obtained by fitting to the data in the $0.8<z<1.4$ panel (corresponding to roughly
the half-way point in our redshift range). At the bottom of each panel
the residual obtained by removing this $0.8<z<1.4$ linear relation is shown. The residuals are flat in all panels except possibly in the lowest redshift bin, where we do not have complete mass coverage. This suggests that for galaxies with masses greater than $10^{10} M_\sun$
the slope of the size-mass relationship remains constant at all redshifts, although its normalization does not. This is in 
good agreement with the findings of \citet{Damjanov2009}, who
reported that the slope of the relation between size $R_e$ and stellar mass $M_*$ of massive quiescent galaxies stays constant, while its zero point smoothly evolves towards lower half-light radii with increasing redshift. 

It is important to consider whether different survey strategies used to obtain the data in Figure ~\ref{f1} play an important role in our interpretation of the observations. The galaxy sizes presented in Figures~\ref{f1}~and~\ref{f2} and listed in Table~\ref{tab2} are measured over a wide range of rest-frame wavelengths $(\lambda_\mathrm{rest}=232-982\, \mathrm{nm})$. However, available data suggest that this is not an important source of error. For example, all but three objects from the EGS subsample \citep{Carrasco2010} have reported sizes based on the imaging that spans the range of  $\lambda_\mathrm{rest}=300-700\, \mathrm{nm}$, where half-light radii show weak dependence on wavelength \citep{Cassata2010}.  Furthermore, the GDDS objects with available NICMOS F160W and ACS F814W images have very similar sizes in both bands, as noted in \S~\ref{data}.  To further investigate possible biases, different selection criteria used to construct the compiled samples are shown in Table~\ref{tab1} coded by font style. Figure~\ref{f2} presents the same data shown in Figure ~\ref{f1}, but with symbols colors keyed to the selection criteria used to define the various surveys. In most of the listed surveys quiescent galaxies have been selected based either on their ultraviolet (UV) absorption spectral features (red points in Figure~\ref{f2}) or on their passive colors (blue points in Figure~\ref{f2}). Four out of 17 subsamples (containing 32\% of all objects) are based on the morphological selection of spheroid-like systems (green points in Figure~\ref{f2}). No trends with selection strategy are seen. 

Perhaps the strongest bias in our sample originates in the spectroscopic selection of passive galaxies at $z>1.5$, since these objects need to be bright enough to be detected in the rest frame UV. Our sample contains 38 objects  (less than 10\% ) in that redshift range and for the high-$z$ surveys in the sample with known spectroscopic completeness level it varies from $\sim50\%$~(GMASS) to $80-90\%$ (GDDS, GS/ACS). Although this may affect the slopes of the size-mass relation in the last two panels of Figures~\ref{f1}~and~\ref{f2}, our main conclusion presented in \S~\ref{size-z} will not be altered since it is heavily based on the lower redshift bins where our selection of galaxies with known $z_{spec}$ is far less biased.

\section{The size growth of quiescent galaxies}\label{size-z}

\subsection{The size-redshift relation}

\begin{figure*}[htp!]
\begin{center}
\vspace {1cm}
\includegraphics[scale=.6,angle=90]{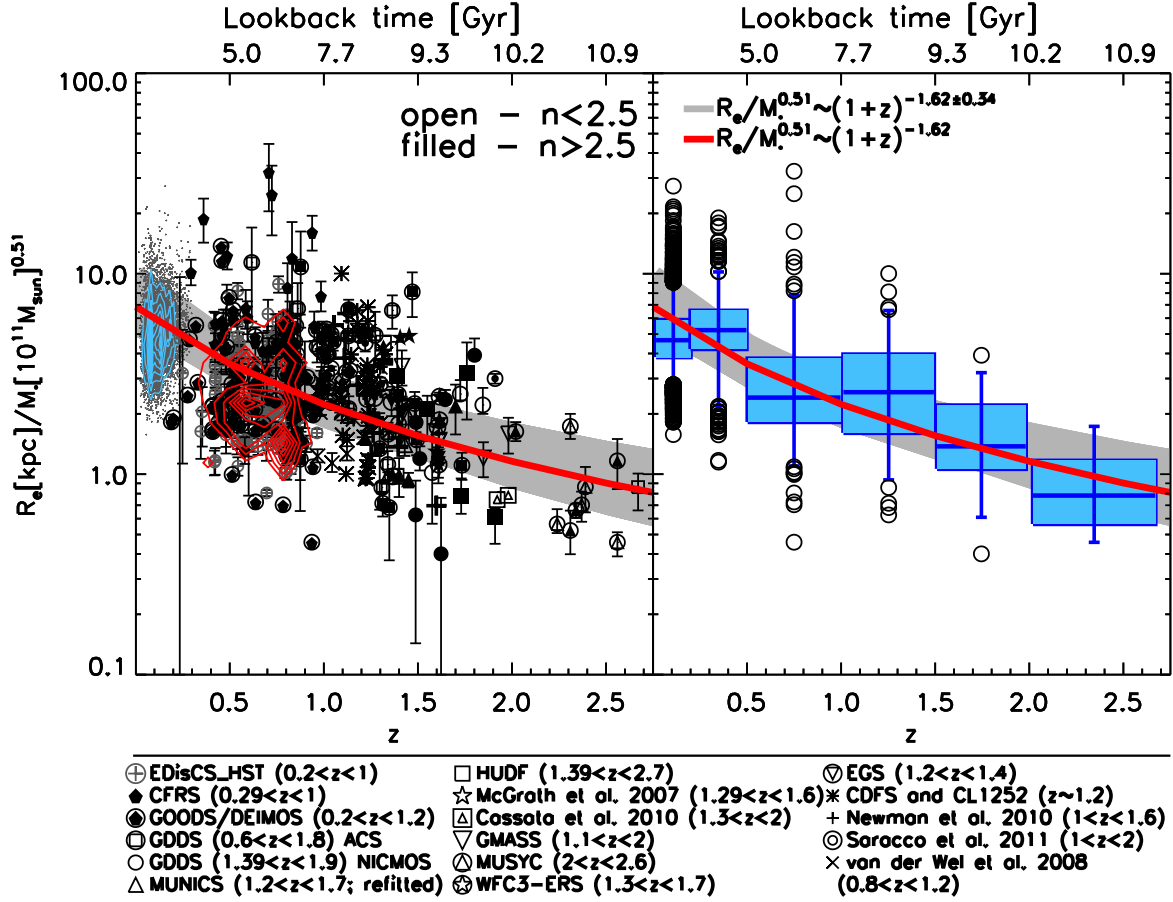}
\vspace {1cm}
\caption{Size evolution of massive quiescent galaxies as a function of redshift. The y-axis represents the effective radius divided by $M_*^{\alpha}$, where $M_*$ is the stellar mass of a galaxy and $\alpha=0.51$ is the slope of the size-mass relation shown in Figure~\ref{f1}. \it{Left:}\rm \ Each symbol type corresponds to a different survey, while blue (red) contours denote the regions of constant density of $z\sim0$ ($0.2<z\lesssim0.9$) galaxies in size-redshift  parameter space. \it{Right:}\rm \ The box-and-whisker diagram for $R_e/M_*^{0.51}$ divided into six redshifts bins. The red line and the grey shaded area in both panels show the best fit to the median redshift points and the $\pm 1 \sigma$~errors of the best relation, respectively. \it{Bottom:}\rm \ The list of spectroscopic surveys included in the presented sample. See text for details.\label{f3}}
\end{center}
\end{figure*}

An even clearer picture of the size evolution of massive quiescent galaxies can be obtained by normalizing out the trends with stellar mass. 
Figure~\ref{f3} shows a plot of size versus redshift in which we have
used the slope $\alpha=0.51$ of the $R_e\propto M_*^{\alpha}$ relation  to normalize the sizes
in order to remove the trend with stellar mass. The full distribution of data is shown in the left-hand panel, while the right-hand panel shows the corresponding `box and whisker' plot\footnote{The top and bottom of the box show the 25th and 75th percentile of the distribution. The horizontal line bisecting the box  is the 50th percentile (the median). The top and bottom of the error bars correspond to the 9th and 91st percentile.
Circles are outliers.}.  A legend mapping data points to individual surveys is provided in the bottom section of Figure~\ref{f3}.

It is interesting to consider whether a smooth function fits the data shown in Figure~\ref{f3}, but a straightforward fit to all the data points would be quite biased. The major portion of the $z>0$ objects $(63\%)$ presented in this paper lies in the redshift range $0.2<z<1$ . On the other hand the local data outnumbers the high-redshift data by orders of magnitude and a straightforward unweighted fit to all the data points clearly places unfair emphasis on fitting the $z=0$  galaxies. Furthermore, the size measurements of the brightest and most massive galaxies in the SDSS sample are affected by the uncertainties in the estimated background sky level producing a steeper slope of the size-mass relation observed locally \citep{Guo2009}. Therefore, as a first (fairly robust) step toward understanding the trends with redshift, we have instead chosen to  calculate the best fit obtained by fitting the median values in the six redshift bins, i.e. giving each redshift range equal weight. This results in $R_e/M_*^{0.51}\propto(1+z)^{-1.62\pm0.34}$ (with the range of 1~$\sigma$ errors obtained by using the bootstrap resampling method). This fit is shown in red in Figure~\ref{f3}, with the corresponding uncertainty shown as a gray band. We emphasize that none of the main conclusions of this paper depend on the specific parametric form represented by this fit.

\subsection{Continuous size evolution with redshift}

The overall conclusion from our analysis is that the median size of massive early-type galaxies 
is continuously growing from $z\sim2.5$ to $z\sim0$. This seems to be in disagreement with some previously reported results showing that a) the size evolution occurs rapidly at $z\gtrsim1$ and becomes negligible at $z<1$ \citep{Fan2010,Valentinuzzi2010,Maier2009} or b) there is no strong evidence for size growth from $z=2$ to $z=0$ \citep{Saracco2010a}. This apparent discrepancy might be the result of earlier studies being
based on samples spanning a limited redshift range, or
which combine spectroscopic and photometric redshift samples,
or which group passively evolving and star-forming objects together, or which contain small number of objects (four things we have tried to avoid doing in the
present paper). On the other hand, perhaps it will eventually prove interesting 
to group {\em some} star-forming objects
with quiescent galaxies at a range of redshifts, since the form of \it{continuous}\rm \ size evolution we obtain for our spectroscopic sample of massive quiescent galaxies is in good agreement with the somewhat shallower size-redshift relation found for UV-bright and submillimetre galaxies in GOODS-North field with secure spectroscopic redshifts over the $z=0.6-3.5$~range \citep[$R_e\sim(1+z)^{1.11\pm0.13}$, ][]{Mosleh2011}. This unexpected concordance hints at a possible evolutionary connection between extreme star-forming and passively evolving galaxies.

Figure~\ref{f3} highlights the main point of our analysis: size growth is both continuous and gradual, at least for the large sample of quiescent objects with spectroscopic redshifts as a whole. It is interesting to compare our results with the ones based on large photometric surveys. Recently, \citet{Williams2010} have performed structural analysis of $\sim3\times 10^4$ star-forming and passively evolving galaxies in the redshift range $z=0.5-2$ from the UKIDSS Ultra-Deep Survey. In addition to the uncertainties introduced by photometric redshifts, the individual size measurements are largely affected by the use of ground based imaging in this survey. Nevertheless, their simulations and empirical tests show that the data provides robust estimates of the \it{average}\rm \ sizes of a large galaxy sample down to $\sim1$~kpc radii. These authors also find a smooth evolution of half-light radii with time for both quiescent and star-forming galaxies described by power laws $(1+z)^\alpha$ with similar exponents that depend on the stellar mass of galaxies and range from $\alpha=-0.75\pm0.10$ for stellar masses $M_*=10^6-10^8\, \mathrm{M}_{\sun}$ to $\alpha=-1.30\pm0.10$ for $M_*>10^{11}\, \mathrm{M}_{\sun}$. A similar trend with mass, i.e., the more prominent size evolution of the most massive quiescent galaxies, has also been found using a small spectroscopic sample of 17 objects in the redshift range $z=1.1-1.6$ \citep{Newman2010}  and at $1\lesssim z\lesssim2.5$ based on a predominantly photometric samples \citep{Ryan2010}.  On the other hand, in a spectroscopic sample of 62 quiescent galaxies at $z=1-2$ with  $M_*=10^{11}-10^{12}\, \mathrm{M}_{\sun}$ the fraction of compact objects does not depend on their mass \citep{Saracco2010a}. All above listed spectroscopic samples are included in our analysis. 

\section{Summary and conclusions}

We have analyzed the size growth of 465 early-type
galaxies taken from 17 spectroscopic surveys
spanning the redshift range $0.2<z<2.7$.
The size evolution of passively evolving galaxies
is continuous and gradual over this redshift range. Size growth appears to be 
independent of stellar mass. Galactic half-light radius
scales with redshift as $R_e/M_*^{0.51}\propto(1+z)^{-1.62\pm0.34}$. Although surveys at higher $z$ are less sensitive to
lower surface brightness galaxies and thus tend to reduce the slopes of the size-redshift relation, based on the lower $z$ distribution this is not expected to be a large effect.
Our resulting power law quantifying smooth size evolution is comparable to the $Re\sim(1+z)^\alpha$ 
relation for massive $(M_*>6.3\times 10^{10}\, \mathrm{M}_\sun)$ quiescent galaxies with the exponent $\alpha=-1.44$ determined by frequent minor mergers at $z=0-2$ in recent cosmological simulations \citep{Oser2011}. 
However, these simplified simulations neither include strong supernova-driven winds nor AGN feedback. Any mechanism proposed to explain size evolution will have to take into account the fact that size growth
is a continuous process that has been occurring more-or-less smoothly and gradually over the last 10 Gyr. 

We thank the referee for a positive feedback. I. D. and R. G. A. acknowledge the financial support provided by the NSERC.


\end{document}